# Creation of atomic coherent superpositions in manifolds of levels without multi-photon resonance

Yueping Niu, Shangqing Gong, Ruxin Li, and Shiqi Jin

Key Laboratory for High Intensity Optics, Shanghai Institute of Optics and Fine Mechanics, Shanghai 201800, P.R. China

We propose a scheme for creating atomic coherent superpositions via stimulated Raman adiabatic passage in a $\Lambda$-type system where the final state has twofold levels. In the employ of a control field, the presence of double dark states leads to two arbitrary coherent superposition states with equal amplitude but inverse relative phase without the condition of multi-photon resonance. In particular, two orthogonal maximal coherent superposition states are created when the control field is resonant with the transition of the twofold levels. This scheme can also be extended to manifold $\Lambda$-type systems.

Coherent superpositions of atomic or molecular states play a crucial role in contemporary quantum physics. Phenomena associated with coherent superpositions have attracted considerable attention and offer a variety of interesting and potential applications, such as electromagnetically induced transparency and coherent population trapping [1, 2], chemical reaction dynamics [3], quantum information and computing [4-6], quantum entanglement [7], and so on. Atoms or molecules prepared in specified quantum state can be realized by technique of $\pi$-pulse, stimulated Raman adiabatic passage [8, 9], chirped pulse [10, 11], etc. Although the $\pi$-pulse technique can prepare atoms in a particular state, this method is not robust because it is very sensitive to small variations of pulse areas. In contrast, the technique of stimulated Raman adiabatic passage (STIRAP) allows a complete population transfer from an initial single state to the target single state robustly. In addition to straightforward population transfer to a single quantum state, the STIRAP technique has widely been applied to create coherent superposition states. For instance, fractional STIRAP [12] has been introduced to construct coherent superposition states, whose composition is determined by the intensity ratio of the pump and Stokes pulses. Because of frequency and intensity fluctuations, however, it is difficult to maintain a constant ratio at all times. Moreover, a tripod linkage has been used for realization of superposition states and measurement of a coherent superposition state with equal amplitudes in metastable neon [13, 14]. Nevertheless, in all the above schemes, the pump and Stokes field must remain two-photon resonance so as to eliminate the excited state from the transfer process and keep high transfer efficiency. On the other hand, in manifold level schemes, chirped adiabatic passage has been applied to prepare population in a superposition state of the final manifold levels [15]. Because of using chirped pulses, however, careful control of chirp rate is required and analytic solutions are difficult to achieve.

The existence of dark state is the basis of electromagnetically induced transparency, coherent population trapping, adiabatic population transfer, etc. In manifold [15, 16] level systems, double dark states present and their interaction has been studied by Lukin *et al.* [16]. Motivated by Ref. [15, 16], in this paper, a $\Lambda$-type system where the final state has a structure of twofold closely spaced levels is introduced to create atomic coherent superposition states. The application of a control field makes condition of multi-photon resonance unnecessary. Because of the existence of double dark states, two arbitrary coherent superposition states with equal amplitude but inverse relative phase are constructed analytically. In the case of the control field resonant with the transition of the two closely spaced levels, two orthogonal maximal coherent superposition states are created simply.

Consider a four level $\Lambda$-type system as shown in Fig. 1. The twofold levels, labeled as 3 and 4, are coupled by a constant control field $\Omega_c$, which can be a microwave or quasistatic field. Level 1 and 2, 2 and 3 are coupled by the pump laser pulse $\Omega_p(t)$ and Stokes laser pulse $\Omega_s(t)$, respectively. The wave functions of the bare state are denoted by |1>, |2>, |3> and |4>. Then the time evolution of the probability amplitudes $C(t) = [C_1(t), C_2(t), C_3(t), C_4(t)]^T$ of the four states is described by the Schrödinger equation:

$$i\hbar \frac{d}{dt} C(t) = H(t) C(t), \tag{1}$$

where the Hamiltonian of this system in the rotating-wave approximation is

$$H = \hbar \begin{bmatrix} 0 & \Omega_p(t) & 0 & 0 \\ \Omega_p(t) & -\Delta_1 & \Omega_s(t) & 0 \\ 0 & \Omega_s(t) & -\Delta & \Omega_c \\ 0 & 0 & \Omega_c & -(\Delta - \Delta_3) \end{bmatrix}. \tag{2}$$

Here, $\Delta_1 = \omega_2 - \omega_1 - \omega_p$, $\Delta_2 = \omega_2 - \omega_3 - \omega_s$ and $\Delta_3 = \omega_3 - \omega_4 - \omega_c$ denote single-photon detunings. $\Delta = \Delta_1 - \Delta_2$ means two-photon detuning. $\hbar\omega_i (i = 1-4)$ describes the energy of level $|i>$ and $\omega_j (j = p, s, c)$ is the carrier frequency of laser $j$. For simplicity, the above detunings are assumed time-independent. Simple calculations show that one of the eigenvalues of Eq. (2) will be equal to zero if

$$\Delta_\pm = \frac{\Delta_3 \pm \sqrt{\Delta_3^2 + 4\Omega_c^2}}{2}, \tag{3}$$

is satisfied. Then the corresponding eigenvector can be introduced as

$$|\psi_0>_\pm = \cos\vartheta |1> - \sin\vartheta (\cos\phi |3> \pm \sin\phi |4>), \tag{4}$$

with the mixing angle $\vartheta$ and $\phi$ defined to be

$$\tan\vartheta(t) = \frac{\Omega_p(t)}{\Omega_s(t)} \sqrt{1 + \frac{\Delta}{\Delta - \Delta_3}}; \quad \tan\phi = \frac{\Omega_c}{|\Delta - \Delta_3|}. \tag{5}$$

Here, the mixing angle $\vartheta$ is similar to the conventional one defined in normal $\Lambda$-type system but is modified by a factor $\alpha = \sqrt{1 + \Delta/(\Delta - \Delta_3)}$. Certainly, since $\alpha$ is time-independent, it will not alter the time evolution of the mixing angle $\vartheta$. Meanwhile, $\phi$ defined here is an additional mixing angle related to detunings and the control field. It is obvious that if the control field is removed, the adiabatic state will be rewritten as $|\psi_0> = \cos\vartheta |1> - \sin\vartheta |3>$ and thus our system evolves into the normal $\Lambda$-type case.

From Eq. (4), one can see that adiabatic state $|\psi_0>_\pm$ associated with the null eigenvalue has no component of the excited state |2>, and hence is immune to the specific properties of state |2>, principally the possible spontaneous emission to other states. So it is a dark state. Moreover, because the coupling of the state $|\psi_0>$ to the other eigenstates is negligible in the adiabatic limit, we do not show the other three eigenstates here.

From Eqs. (3) and (4), we can see clearly that corresponding to $\Delta_+$ and $\Delta_-$, there exists two dark states

$|\psi_0>_+$ and $|\psi_0>_-$. For delayed counterintuitive pump and Stokes pulses (the Stokes pulse precedes the pump pulse), the relations $0 \xleftarrow{t \to -\infty} \Omega_p(t)/\Omega_s(t) \xrightarrow{t \to \infty} \infty$ apply. As time progresses from $-\infty$ to $\infty$, the mixing angle $\vartheta$ rises from 0 to $\pi/2$. Consequently, the adiabatic state $|\psi_0>_\pm$ starts in the bare state |1> will end in coherent superposition state $|\Phi>_+ = \cos\phi$ |3> + $\sin\phi$ |4> or $|\Phi>_- = \cos\phi$ |3> - $\sin\phi$ |4>. These two superposition states have equal amplitude but inverse relative phase. Suitable manipulation of detunings and control field permits any value of $\phi$ we desired. Therefore, at the end of pump pulse, either a single state or arbitrary superposition of |3> and |4> can be realized. It is important to note that this process needs neither two-photon nor tri-photon resonance. Apparently, it is the control field that compensates the multi-photon detunings and therefore the absolute transfer efficiency is kept without condition of multi-photon resonance.

It is noteworthy that when the control field is on-resonance, i.e. $\Delta_3 = 0$, the ultimate superposition state $|\Phi>_\pm$ has a form of $|\Phi>_+ = 1/\sqrt{2}$ (|3>+|4>) for $\Delta_+ = \Omega_c$ and $|\Phi>_- = 1/\sqrt{2}$ (|3>-|4>) for $\Delta_- = -\Omega_c$. A special feature here is the target state now is a maximal coherent superposition of the twofold levels and $|\Phi>_+$ is orthogonal to $|\Phi>_-$. They are possible states of the quantum bit and hence have essential applications in quantum phase gate [5, 17].

To illustrate the preceding analytic solutions, we present numerical simulations in the case of Gaussian pulses with the same duration $T$ for the pump and Stokes lasers. $\Omega_p(t) = \Omega_p \exp[-(t-\tau)^2/T^2]$, $\Omega_s(t) = \Omega_s \exp[-(t+\tau)^2/T^2]$. The Stokes pulse precedes the pump pulse and the delay time between them is $2\tau$. $\Omega_s$ and $\Omega_p$ are peak values of the two pulses. During our numerical simulations, Eq. (3) and $\Omega_s(\Omega_p)T \gg 1$ (adiabatic criterion) is fulfilled. Without loss of generality, the system is assumed to be initially in state |1>, i.e. $C_1(-\infty) = 1$, $C_{2,3,4}(-\infty) = 0$.

By solving Schrödinger Eq. (1), we show the time evolutions of populations. Here $T = 5T_0$ and $\tau = 2.5T_0$. We scale all parameters in terms of $T_0$. As Fig. 2(a) reveals, under the circumstances of $\Delta_3 = 0$, the twofold levels |3> and |4> obtain equal population at the end of the pump pulse while levels |1> and |2> are empty. Figures 2(b) and 2(c) show time evolutions of probability amplitudes $C_3$ and $C_4$. After all the interactions are over, $C_3 = C_4$ when $\Delta = \Omega_c$ while $C_3 = -C_4$ when $\Delta = -\Omega_c$. As discussed above, these two cases correspond to two orthogonal coherent superposition states $|\Phi>_+$ and $|\Phi>_-$. We find that under the condition of adiabatic evolution, resonant control field ($\Delta_3 = 0$) guarantees the creation of two precisely orthogonal maximal coherent superposition states $|\Phi>_+$ and $|\Phi>_-$.

From Eq. (4) one can derive the final population probability ratio of level |3> to |4>: $R = (\cos\phi/\sin\phi)^2$. Figure 3 illustrates some examples. As is well known, the STIRAP technique applied in normal $\Lambda$-type system can transfer population from the initial state to the final one completely but two-photon resonance must be met critically. Otherwise, the excited level will be populated during the transfer process and thus its decay will deteriorate the transfer efficiency sharply. However, in the present case, an efficient population transfer without any population on the excited level can be realized although two-photon detuning $\Delta \neq 0$. Therefore, we can say that preselected atomic single or superposition quantum state can be realized through proper choice of parameters according to the analytic expression of the dark state $|\psi_0>_\pm$ without any resonant conditions to be satisfied.

As for the adiabatic criterion, we display the other three non-zero eigenvalues (solid line) and the rate of change in the mixing angle $\dot{\vartheta}$ (dotted line) numerically. From Fig. 4, one can see that $\dot{\vartheta}$ is so small compared with separations of the eigenvalues. Hence, nonadiabatic coupling between the eigenstates is negligible and adiabatic condition could be fulfilled.

Additionally, this $\Lambda$-type system with twofold can be extended to manifold $\Lambda$-type systems, just as shown in Fig. 5(a). Here we put forward a three-fold system depicted in Fig. 5(b). Two resonant control fields $\Omega_c$ and $\Omega_d$ are applied and the other parameters have the same meaning as in the above twofold system. Provided that $\Delta = \pm\sqrt{\Omega_c^2 + \Omega_d^2}$, two dark states will come into existence. Set $\varphi$ to be $\tan^{-1}(\Omega_s\Omega_c/\sqrt{2}\Delta\Omega_p)$, then the adiabatic state $|\psi>_\pm = \sin\varphi\,|1> - \cos\varphi\,(\Omega_c|3> \pm |\Delta||4> + \Omega_d|5>)/\sqrt{\Delta^2 + \Omega_c^2 + \Omega_d^2}$ can be indicated. By adjusting the field parameters and the detunings, desired single quantum state or superposition state can be constructed. Figure 5(c) presents an example of numerical simulation. In a similar manner, N-component coherent superposition states can be created in the manifold system under proper choice of parameters.

In summary, we have explored both analytically and numerically the creation of superpositions in the $\Lambda$-type system where the final state has two closely spaced levels. As long as Eq. (3) is satisfied, two arbitrary coherent superposition states with equal amplitude but inverse relative phase can be prepared without multi-photon resonance. Special attention has been paid to the case of the control field resonant with the corresponding transition frequency. It can be seen that under this condition, $|\Phi>_+$ and $|\Phi>_-$ are orthogonal maximal coherent superposition states. Proper choice of parameters provides more freedom in manipulating the atomic coherent superposition states. To be brief, our scheme, which is usually feasible in a realistic experimental situation, such as [18], demonstrates a possibility to create preselected atomic coherent superposition states.

This work is supported by the National Natural Sciences Foundation of China (No. 10234030). Shangqing Gong thanks K. Bergmann, M. Fleischhauer and B. W. Shore for the hospitality during his stay in Kaiserslautern.

Captions:

FIG. 1. Schematic diagram of a $\Lambda$-type system where the final state is a twofold of closely spaced levels.

FIG. 2. The time evolutions of (a) populations $P_i = |C_i|^2$ ($i$=1-4). (b) and (c) amplitudes $C_3$ and $C_4$. Parameters are $\Omega_p = \Omega_s = 4.0$, $\Omega_c = 2.5$, $\Delta_1 = 3.5$, and $\Delta_3 = 0$. (a) $\Delta_2 = 1.0$ or 6.0. (b) $\Delta_2 = 1.0$. (c) $\Delta_2 = 6.0$.

FIG. 3. The time evolutions of populations $P_i = |C_i|^2$ ($i$=1-4). Parameters are (a) $\Omega_p = \Omega_s = 4.0$, $\Omega_c = 1.5$, $\Delta_1 = 2.0$, and $\Delta_2 = 1.0$. (b) $\Omega_p = \Omega_s = 4.0$, $\Omega_c = 3.0$, $\Delta_1 = 0.0$, and $\Delta_2 = 0.2$. (c) $\Omega_p = 2.0$, $\Omega_s = 9.0$, $\Omega_c = 1.7$, $\Delta_1 = 11$, and $\Delta_2 = 1.0$.

FIG. 4. The three non-zero eigenvalues (solid line) and the rate of change in the mixing angle $\dot{\vartheta}$ (dotted line). Parameters are $\Omega_p = \Omega_s = 4.0$, $\Omega_c = 2.5$, $\Delta_1 = 3.5$, $\Delta_2 = 6.0$, and $\Delta_3 = 0.0$.

FIG. 5. (a) and (b) The extended manifold level systems. (c) The time evolutions of populations corresponding to threefold level system as shown in (b). Parameters are $\Omega_p = \Omega_s = 4.0$, $\Omega_c = 3.0$, $\Omega_d = 4.0$, $\Delta_1 = 4.0$, $\Delta_2 = 5.0$, and $\Delta_3 = -1.0$.

Figures:

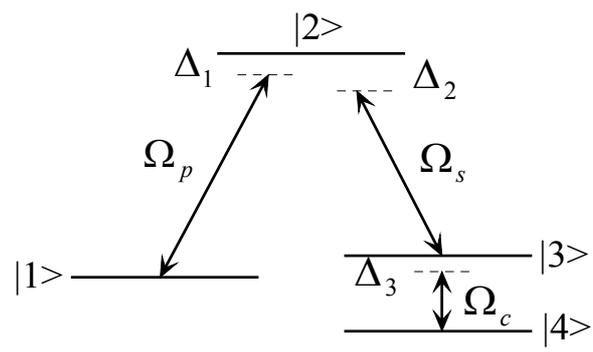

FIG. 1, Niu *et al.*

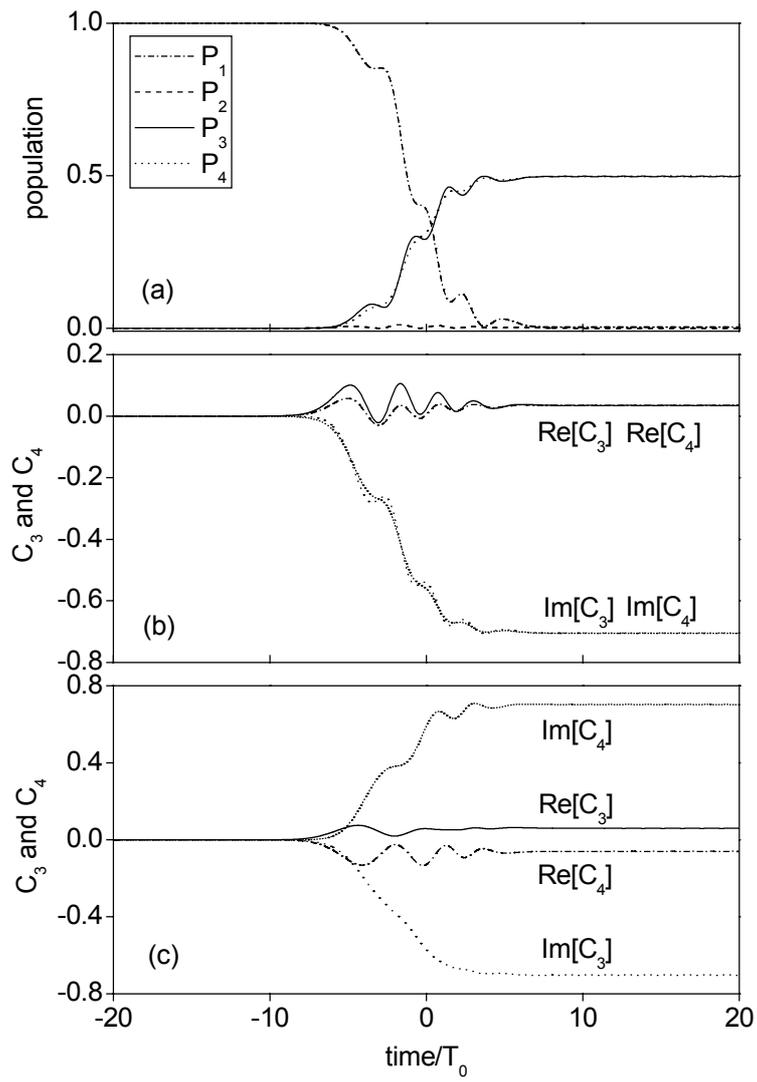

FIG. 2, Niu *et al*.

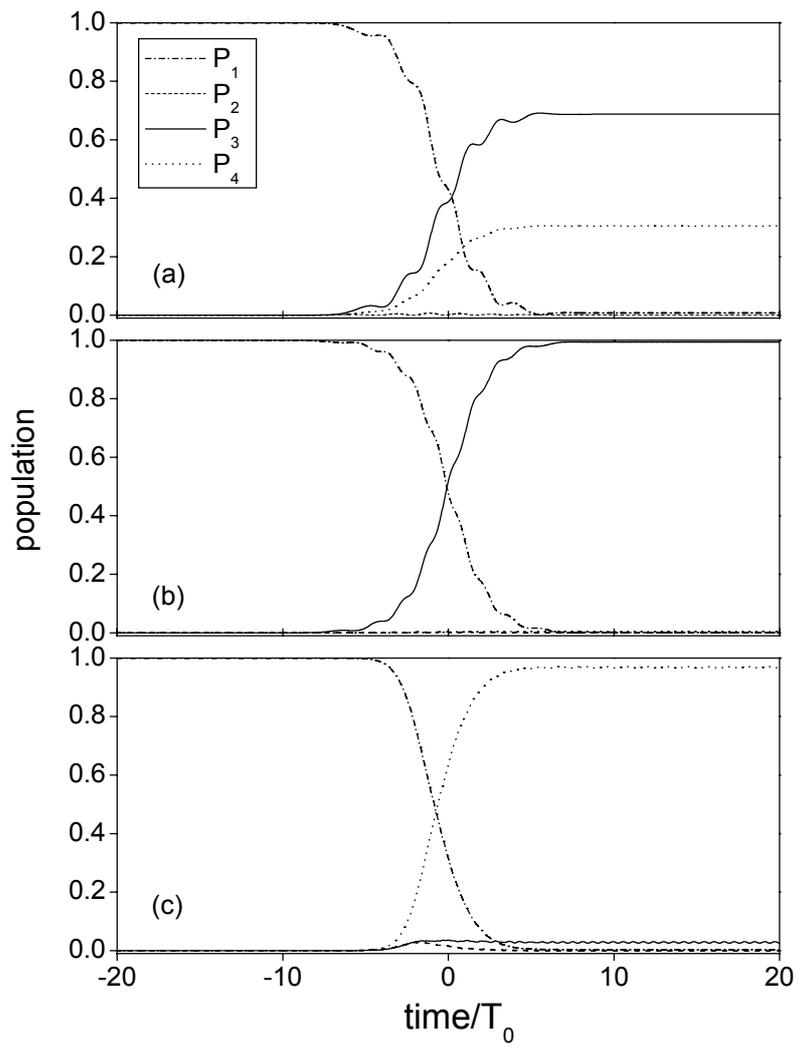

FIG. 3, Niu *et al*.

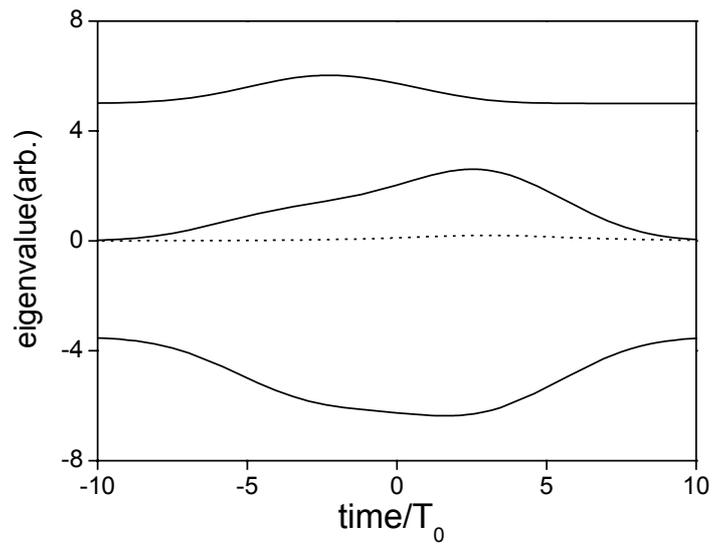

FIG. 4, Niu *et al*.

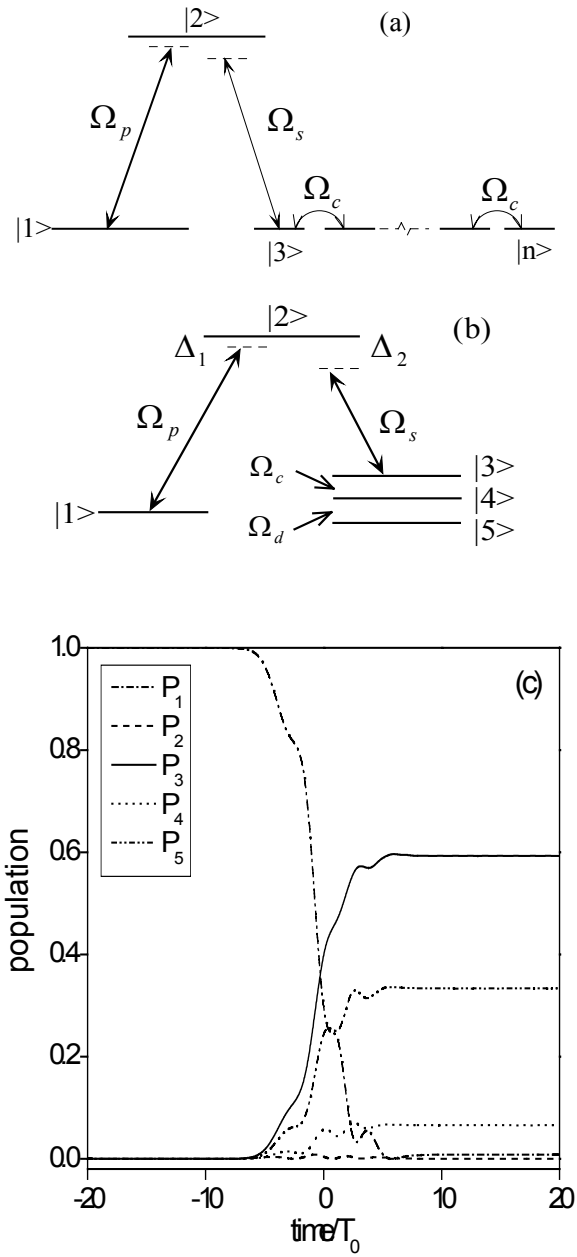

FIG. 5, Niu *et al*.